# Nanostructural Superconducting Materials for Fault Current Limiters and Cryogenic Electrical Machines

Prikhna T.A.[a], Gawalek W.[b], Savchuk Ya.M.[a], Sergienko N.V.[a], Moshchil V.E.[a], Sokolovsky V.[c], Vajda J.[d], Tkach V.N.[a], Karau F.[e], Weber H.[f], Eisterer, M.[f], Juolain A.[g], Rabier J.[g], Chaud X.[h], Wendt M.[b] Dellith J.[b], Danilenko N. I.[k] Habisreuther T.[b], Dub S.N.[a], Meerovich V.[c], Litzkendorf D.[b], Nagorny P.A.[a], L.K. Kovalev[l], Schmidt Ch.[b], Melnikov V. S.[a], Shapovalov A.P.[a], Kozyrev A.V.[a], Sverdun V.B.[a], Kosa J.[d], Vlasenko A. V.[a].

[a] Institute for Superhard Materials of the National Academy of Sciences of Ukraine, 2 Avtozavodskaya Street, Kiev, 04074, Ukraine, prikhna@mail.ru, prikhna@iptelecom.net.ua

[b] Institut für Photonische Technologien, Albert-Einstein-Strasse 9, Jena, 07745, Germany, gawalek@ipht-jena.de,

[c] Ben-Gurion University of the Negev, P.O.B. 653, Beer-Sheva 84105 Israel, sokolovv@bgu.ac.il

[d] Budapest University of Technology and Economics, Budapest, Hungary 1111 Budapest, Egry Jozsef u. 18. Hungary, vajda@supertech.vgt.bme.hu

[e] H.C. Starck GmbH, Goslar 38642, Germany

[f] Atomic Institute of the Austrian Universities, 1020 Vienna, Austria, weber@ati.ac.at

[g] Universite de Poitiers, CNRS/ Laboratoire PHYMAT, UMR 6630 CNRS-Universite de Poitiers SP2MI, BP 30179, F-86962 Chasseneuil Futuroscope Cedex, France, jacques.rabier@univ-poitiers.fr, anne.joulain@univ-poitiers.fr

[h] CNRS/CRETA Grenoble, 38042, CEDEX 9, France, xavier.chaud@grenoble.cnrs.fr

[l] Institute for Problems of Materialscience of the National Academy of Sciences of Ukraine,3, Krzhyzhanivsky Str., Kiev 03142, Ukraine

[k] Moscow State Aviation Institute (Technical University) Volokolamskoe Shosse, 4, 125993 Moscow, Russia

Materials of the Y-Ba-Cu-O (melt-textured $YBa_2Cu_3O_{7-\delta}$-based materials or MT-YBCO) and Mg-B-O ($MgB_2$-based materials) systems with high superconducting performance, which can be attained due to the formation of regularly distributed nanostructural defects and inhomogenities in their structure can be effectively used in cryogenic technique, in particular in fault current limiters and electrical machines (electromotors, generators, pumps for liquid gases, etc.).

The developed processes of high-temperature (900-800 $^{o}C$) oxygenation under elevated pressure (16 MPa) of MT-YBCO and high-pressure (2GPa) synthesis of $MgB_2$-based materials allowed us to attain high SC (critical current densities, upper critical fields, fields of irreversibility, trapped magnetic fields) and mechanical (hardness, fracture toughness, Young modulus) characteristics. It has been shown that the effect of materials properties improvement in the case of MT-YBCO was attained due to the formation of high twin density (20-22 $\mu m^{-1}$), prevention of macrockracking and reduction (by a factor of 4.5) of microcrack density, and in the case of $MgB_2$-based materials due to the formation of oxygen-enriched as compared to the matrix phase fine-dispersed Mg-B-O inhomogenities as well as inclusions of higher borides with near-$MgB_{12}$ stoichiometry in the Mg-B-O matrix (with 15–37 nm average grain sizes). The possibility is shown to obtain the rather high $T_c$ (37 K) and critical current densities in materials with $MgB_{12}$ matrix (with 95 % of shielding fraction as calculated from the resistant curve).

# Nanostructural Superconducting Materials for Fault Current Limiters and Cryogenic Electrical Machines

**1. Introduction.**

Interest in the superconductive (SC) nanostructural bulk materials of the Y-Ba-Cu-O and Mg-B-O systems is aroused because of their unique superconducting properties (possibility to attain high critical current densities, $j_c$, upper critical fields $H_{C2}$, fields of irreversibility, $H_{irr}$, to trap high magnetic fields, $B_z$) and because of the modern progress directed toward technologies utilizing liquid hydrogen (as a fuel for combustion engines of different types of transport and at the same time cooling agent, etc.). The high thermomechanical characteristics of the materials are also of great importance, because when operated they are permanently subjected to essential stresses of electromagnetic fields and of thermal strains as a result of thermocycling from room temperature down to the working temperatures or due to a release of the Joule heat as a result of SC-normal transition. Thus, for the high functional conformance of SC devices the mechanical characteristics of materials like hardness, $H_v$, fracture toughness, $\kappa_{1C}$, and Young modulus, $E$, should be high enough.

MT-YBCO (melt-textured $YBa_2Cu_3O_{7-\delta}$-based or in short Y123-based) and $MgB_2$-based ceramic materials can be successfully used in SC electromotors, cryogenic pumps, fault current limiters, magnetic bearings, fly-wheel energy storage devices, MAGLEV transport, magnetron-sputtering devices, etc. operating at liquid nitrogen and hydrogen temperatures [1, 2].

Fault current limiters (FCL) can be one of the primary applications of high temperature superconductors in electric power systems. Superconducting FCLs (SC FCLs) have no analogs in traditional electrical engineering and are designed to increase the reliability of power delivery and to protect electric power equipment against the electrodynamic and thermal stresses [3]. From the technical point of view the use of a bulk SC material is most reasonable in inductive (transformer) type of SC FCLs. A typical design of an SC inductive fault current limiter is that where the superconducting-normal transition in the cylindrical ring from superconducting material causes the increase of the impedance of the primary windings in the case of overloading. Essentially an SC FCL is a self-triggered, self-restoring (self-regenerated, i.e. it returns again to the initial state when the current surge is decreased) device with a fast response and low losses under normal operation [3]. Cryogenic electromotors (generators, pumps for liquid gas) are characterized by a high energy density at the rotor surface and usually are 7-10 times more compact and of lower weight than the traditional ones. A decrease of the rotor size of a SC-based electromotor leads to a decrease of the inertia moment of the motor, which is of great importance in the case of the operation in the regime of often reverse (for example, to guide the looms or stamping presses, special stands for automobile testing, etc.). It turns out that to use the bulk superconductors in the reluctance type motors is most effective [4, 5].

In the Y123-based superconductors the grain boundaries with the high angles of misorientation (higher than 5 degrees) [6] are a significant obstacle for the superconductive current flow due to a low coherence length of Y123: 0.6-3.1 nm [7], and their presence in the material results in an essential decrease of $j_c$. For the attainment of $j_c$ higher than $10^3$ A/cm$^2$ a quasi-single-crystalline or textured Y123 structure with low misorientation angles between crystallites should be obtained [8] and for such purposes the so called top-seed melt-textured growth can be used. The pinning centers (which promote the $j_c$ increase due to the non-SC vortexes fixed on them) in a Y123-based material can be dislocations, stacking faults, and twins [9] and for reaching high $j_c$ the density of such defects in quasi-single-crystalline or textured Y123 should be so much high that the material can be considered nanostructural. The high density of such structural defects can be formed by choosing the oxygenation conditions [10, 11]. The higher coherent length of MgB$_2$ (1.6-12 nm [12]) as compared to YBa$_2$Cu$_3$O$_{7-\delta}$ (or Y123), gives the possibility to attain high critical current densities, $j_c$, and trapped magnetic fields in the polycrystalline MgB$_2$ - based material, because the grain boundaries are not the obstacles for superconductive current flow as in case of Y123 and can act as pinning centers, so the structure with nanosized grains is preferable for higher $j_c$. Besides, the pinning centers in

$MgB_2$ can be nanosized inclusions of the second phases [13-15]. One can increase the critical current density of the material by chemical alloying [16].

Usually the structure of MT-YBCO (top-seed grown, for example) [17] is constituted from quasi single-crystalline or textured superconducting Y123 matrix growing in the whole volume of the sample with fine-dispersed inclusions of nonsuperconductive, the so-called "green|", $Y_2BaCuO_5$ (Y211) phase. It is considered that the presence of dispersed inclusions of the Y211 phase in the superconductive structure improves the pinning due to the formation of high density of dislocations and stacking faults around Y211 grains [18].

In the present paper the authors-developed methods of high-pressure synthesis of $MgB_2$-based materials and oxygenation of MT-YBCO under elevated pressure at high temperatures, which allows to produce bulk materials with the highest in nowadays SC and mechanical characteristics are discussed as well as the efficiency and perspectives of the application of the materials manufactured by these methods in SC FCLs and electromotors.

## 2. Experimental.

For high-pressure high-temperature (HP-HT) synthesis recessed-anvil type high-pressure apparatuses (HPA) described elsewhere [19] with a working volume up to 330 $cm^3$ were used. During the synthesis $MgB_2$-based materials were in contact with precompacted hexagonal BN powder to prevent their contact with a graphite heater. Using high pressure (2 GPa)–high temperature (600-1050 °C) technique, $MgB_2$ – based blocks up to 62 mm in diameter and 20 mm thick of near theoretical density, with average grain sizes 15–37 nm [14] and high SC characteristics [20] can be produced as well as ring-shaped products and parts of complicated configuration from these blocks.

The preparation of MT–YBCO ceramics includes two main stages: 1) the formation (using a seed crystal) of a pseudo-single-domain $YBa_2Cu_3O_{7-\delta}$ ($\delta \sim 0.8–1.0$) or Y123 structure with finely dispersed inclusions of the $Y_2BaCuO_5$ (Y211) non-superconducting phase - the so-called melt texturing process, 2) the oxygenation of the $YBa_2Cu_3O_{7-\delta}$ matrix structure of the MT–YBCO ceramics to impart superconducting properties to the material by increasing the oxygen content of the CuO basal planes of the $YBa_2Cu_3O_{7-\delta}$ matrix (in this case, δ decreases to 0.2–0). The melt textured method allows us to produce MT-YBCO high-quality blocks of size up to $45 \times 45 \times 18$ mm$^3$ and $90 \times 45 \times 18$ mm$^3$.

In the present study, two types of starting MT-YBCO have been used (Fig. 1). Type 1 is a traditional bulk MT-YBCO produced from a mixture of a commercial $YBa_2Cu_3O_{7-\delta}$ powder (Solvay) with $Y_2O_3$ and $CeO_2$ powders taken in the standard ratio of $Y_{1.5}Ba_2Cu_3O_{7-\delta}$+1%$CeO_2$, the texturing process using seed crystals being performed in air [21]. Type 2 is thin-walled MT-YBCO (with drilled holes having the walls between each other of about 3 mm in order to reduce the depth on which oxygen should penetrate during oxygenation) that has been manufactured [22] from the mixture of 70 wt% of $YBa_2Cu_3O_{7-\delta}$, 30 wt% of $Y_2BaCuO_5$, to which 0.15 wt% of $PtO_2$ has been added. The cooling of Type 2 MT-YBCO after the formation of the Y123 texture using a seed crystal starts from 980 °C in a low-oxygen atmosphere (below 0.5 kPa) and 0.1 MPa nitrogen atmosphere instead of in air in order to prevent the material cracking. The starting MT-YBCO samples of Types 1 and 2 had a tetragonal textured structure of the $YBa_2Cu_3O_{7-\delta}$ matrix with δ≈0.9-0.8.

The oxygenation to increase the oxygen content in $YBa_2Cu_3O_{7-\delta}$ structure of MT-YBCO to δ≈0.2-0.0 in order to impart the superconductive properties to the material has been performed in the separate processes (at atmospheric pressure under flowing oxygen and under elevated oxygen pressure up to 16 MPa). The oxygenation under the isostatic oxygen pressure was performed in a cylindrical stainless steel apparatus (gasostat), which can be filled with oxygen up to a pressure of 16 MPa and heated at this pressure to 800 °C. The oxygenation was conducted as follows: after heating from 20 °C to 700 °C at a rate of 46° per hour the nitrogen atmosphere (1 bar pressure) was replaced by the oxygen one (the gases flowing through the vessel) in accordance with the exponential law. When the temperature was increased to 800 °C, the samples were in the pure oxygen atmosphere (under 1 bar pressure). Then maintaining the temperature at about 800 °C we increased the oxygen pressure (according to the exponential law) up to 16 MPa, under these conditions the samples were kept for 20 h, then the heating was switched off or temperature was decreased at a rate of 60 K/h and the samples were allowed to cool to room temperature. All in all, the samples were held at 800 °C for 60 hours. Such a complicated procedure of the so-called progressing oxygenation was chosen to reduce cracking (to obtain the material with high $j_c$) by keeping the equilibrium oxygen content of the Y123 phase. To study the oxygenation process, large MT-YBCO blocks were cut into rectangular parallelepipeds of size about 2.5-3×2.5-3×6.5-8.5 mm$^3$ (that could be inserted into the magnetometer as a whole) with a diamond wire in such a manner that the longer side of the parallelepiped was approximately parallel to the *c*-axis of the Y123 domain (the orientation of *c*-axis and *ab*-plains of Y123 textured matrix are dependant on the orientation of a seed crystal and can be well seeing in polarized light). The structure of materials was examined using polarized optical, SEM, and TEM (at 200 kV) microscopes as well as X-ray structural and diffraction analyses. For the $YBa_2Cu_3O_{7-\delta}$ phase in the range of $0 \leq \delta \leq 1$ there exists the approximately linear dependence between the *c*-parameter and oxygen content of the phase. We calculated δ in accordance with the established equation δ=60.975 *c* -71.134 [23]. The $j_c$ was estimated from magnetization hysteresis loops obtained on an Oxford Instruments 3001 vibrating sample magnetometer (VSM) using Bean's model. For VSM measurements, we used the whole sample and this size was applied to calculate the $j_c$ -values. The trapped magnetic field distribution over the field-cooled blocks was found using the Hall probe (the distance from the Hall probe to the sample surface was 0.8 mm).

The microhardness, nanohardness and Young modulus were estimated employing a Matsuzawa Mod. MXT-70 microhardness tester, $H_V$ (using a Vickers indenter) and Nano-

Indenter II, $H_B$ (using a Berkovich indenter). The fracture toughness was estimated from the length of the radial cracks emanating from the corners of an indent.

The SEM study of $MgB_2$-based materials were performed using Electron Microprobe analysis (EPMA) by a JEOL JXA 8800 Superprobe and ZEISS EVO 50XVP (resolution of 2 nm at 30 kV), equipped with the INCA 450 energy-dispersion analyzer of X-ray spectra (OXFORD, England) and with the HKL Canell 5 detector of backscattering electrons (OXFORD, England).

### 3. MT-YBCO

From the results shown in figure 2 one can see that high-temperature (800 $^o$C) oxygenation under controllable oxygen pressure (0.5 kPa – 16 MPa) for 80 h allows one to increase the critical current density of MT-YBCO as compared to the long-term 270 h oxygenation under atmospheric oxygen pressure (under the conditions, which are considered to be optimal). Besides, some increase in microhardness and essential increase of fracture toughness (Table 1) have been observed. This increase stems from reducing the macro- and microcracking and increasing the twin density (Fig. 3). The possibility to oxygenate a Y123 structure at high temperature under oxygen pressure (16 MPa) has been demonstrated (which should not be possible under such lower pressure from the point of view of the dependences of partial oxygen pressure given in [24]) as well as the possibility to regulate the amount of microcracks and twins changing the oxygenation temperature [25]. The correlation of the twin and microcrack density with the size and distribution of Y211 inclusions in the Y123 matrix of MT-YBCO has been observed: the more homogeneously distributed but coarser grains of 211 phases induce the higher microcracks and twin density formation (see, for example Figs. 3a-g).

The importance of twins in the Y123 structure for attaining high critical current has been experimentally shown. MT-YBCO with high twin density practically free from dislocations and stacking faults has been obtained (Figs. 3f, g) by oxygenation at high temperature and oxygen pressure and it demonstrates record high values of $j_c$ (Figs. 2a, b). However, our study shows that the detwinned under high pressure but with high dislocation density materials (Figs. 3 h, k) demonstrated essentially lower $j_c$ (Figs.2 a, b). The oxygenation under elevated pressure and high temperature especially improve SC properties of thin-walled ceramics as witnessed by a much higher level and character of distribution of trapped magnetic field in thin-walled MT-YBCO ceramics than that in the bulk one prepared using the same initial powders (Figs. 4). It has been observed that the oxygenation of MT-YBCO under isostatic oxygen pressure (up to 16 MPa) at 800 °C allowed a reduced process time, lower macrocracking, and reduced microcracks. Moreover, higher (the highest in nowadays) critical currents, trapped fields and mechanical characteristics can be attained. At 77 K thin-walled MT-YBCO had a $j_c$ of 85 kA/cm$^2$ in the *ab* plane at 0 T and higher than 10 kA/cm$^2$ in fields up to 5 T; the irreversibility field being 9.8 T. In the c- direction $j_c$ was 34 kA/cm$^2$ in 0 T and higher than 2.5 kA/cm$^2$ in a 10 T field. At 4.9 N-load the microhardness, $H_v$, was 8.7±0.3 GPa in the *ab*-plane and 7.6±0.3 GPa in the *c*-direction. The fracture toughness, $K_{1C}$, was 2.5±0.1 MPa·m$^{0.5}$ (*ab*-plane) and 2.8± 0.24 MPa·m$^{0.5}$ (*c*-direction). Samples with a higher twin density demonstrated a higher $j_c$, especially in the applied magnetic field. The twin density correlates with the sizes and distribution of Y211 grains in Y123. The thin-walled high dense ceramics with homogeneously distributed but coarse Y211 grains demonstrated the highest $j_c$ and contained about 22 twins in 1 μm and were practically free from dislocations and stacking faults. The maximal trapped field of the block of thin-walled ceramic oxygenated at 800 °C and 16 MPa was doubled as compared to that oxygenated at low temperature under ambient pressure.

### 4. Mg-B-O (or MgB$_2$) – based ceramics.

It has been established in our previous study [14] that superconductive properties ($j_c$, in particular) of MgB$_2$ high-pressure high-temperature (HP-HT) synthesized depend on the amount, size and distribution of MgB$_{12}$ inclusions (the finer the inclusions and the larger amount of them, the higher $j_c$). Besides, it has been shown that additions of Ti, Ta and Zr can essentially improve $j_c$ of HP-HT-synthesized materials [20]. On the one hand, they absorb impurity hydrogen and on the other promote an increase in the amount of $MgB_{12}$. The investigation of the oxygen distribution in the materials synthesized and sintered at 2 GPa in the range from 700 to 1100 $^o$C shows that magnesium diboride matrix with near $MgB_2$ stoichiometry contains about 3.5–14 % oxygen, the black inclusions with stoichiometry close to $MgB_{12}$ contains practically no oxygen (0 - 1.5%) [27]. The correlations have not been found between the amount of oxygen in the starting B or $MgB_2$ and its amount in the materials produced as well as between the oxygen content and $j_c$. Here we mean the oxygen content distributed in the materials matrix, but not the oxygen incorporated in MgO (the amount of MgO in all studied materials was rather small and practically no correlations between the MgO amount and manufacturing conditions were found) [27]. We cannot detect $MgB_{12}$ by X-ray. It is known [28] that $MgB_{12}$ reflexes can be absent in the X-ray pattern due to poor diffracted signals because of the low X-ray atomic scattering factor of boron; besides, $MgB_{12}$ inclusions are dispersed in $MgB_2$, the etalon X-ray pattern of $MgB_{12}$ is absent in the database and the literature data are contradictive (some authors mentioned the orthorhombic structure of $MgB_{12}$ and the other ones pointed out that it was rhombohedral or hexagonal) .

In materials synthesized with Ti or SiC at 1050 $^o$C (from amorphous B and Mg taken in the $MgB_2$ ratio with 10% of Ti or SiC) as compared to that synthesized at 800 $^o$C: the segregation of oxygen with the formation of Mg-B-O inclusions (marked "A" in Figs. 5a,b) and a decrease in the oxygen content of the material matrix ("B") leads to the $j_c$ increase (see, for example such structures in Figs. 6 a, b), which may be the result of cleaning of $MgB_2$ grain boundaries.

Materials synthesized from the mixture of Mg and B taken in 1:4, 1:6, 1:7, 1:8, 1:10, 1:12, and 1:20 ratios were superconducting with $T_c$ of about 37 K. The highest $j_c$ ($7·10^4$–$2·10^4$ A/cm$^2$ in zero field at 10 –30 K, respectively, Fig. 7b) was exhibited by the materials having the matrix composition near $MgB_{12}$ stoichiometry (Fig. 7a) prepared from Mg:B=1:8, besides, the estimated from the resistant curve content of a shielding fraction has been about 95% (taken into account the so-called shape corrections), so, the sample contained big volume of SC phase (with

$T_c$ near 37 K). Only very seldom inclusions with near $MgB_2$ composition of 100 nm were observed using high-resolution TEM (JEM-2100F TEM equipped with an Oxford INCA energy detector, the diameter of probe being 0.7 nm) in the material.

Ti, Zr and Ta additions can improve $j_c$ of $MgB_2$-based materials by promoting the higher boride formation via impurity hydrogen absorption, thus preventing $MgH_2$ detrimental for $j_c$ being formed, which possibly increases the $MgB_{12}$ nucleation barrier. SiC (0.2-0.8 μm) addition increases $j_c$ of $MgB_2$, allowing us, for example, to get $j_c=10^6$ A/cm$^2$ at 20 K in the 1T field: pinning is increased by SiC and higher boride grains, and there is no notable interaction between SiC and $MgB_2$. As the synthesis temperature increases from 800 to 1050 °C, Ti and SiC additions may affect the oxygen segregation and formation of Mg-B-O inclusions enriched with oxygen as compared to the amount of oxygen in the $MgB_2$ matrix, which can also promote an increase in pinning.

The hardness of the HPS material (HPS-$MgB_2$ with 10% Ta) measured by a Vickers indenter under a 148.8 N-load is $H_v$= 10.12±0.2 GPa and the fracture toughness under the same load is $K_{1C}$=7.6± 2.0 MPa m$^{0.5}$. The HPS-$MgB_2$ without additions has $H_v$=16.85± 0.74 GPa and $K_{1C}$=4.4±0.14 MPa m$^{05}$ under a load of 4.96 N. Mechanical characteristics of the $MgB_{12}$ phase are higher than that of $MgB_2$ and sapphire $Al_2O_3$ (nanohardness at 50 mN were 32.4 GPa of $MgB_{12}$, 27.3 GPa of $Al_2O_3$, and 17.4 GPa of $MgB_2$; Young moduli were 385±14 of $MgB_{12}$ and 416 ± 22 of $Al_2O_3$; microhardness at 4.9 N were 32 GPa of $MgB_{12}$, 25.1 GPa of $Al_2O_3$, and 16.9 GPa of $MgB_2$).

**5. Application of MT-YBCO and $MgB_2$-based materials in cryogenic electrical machines and in fault current limiters.**

The attained level of SC and mechanical properties using the developed technologies were the highest for the bulk MT-YBCO and MgB$_2$-based materials described in the literature.

In the Budapest University of Technology and Economics the experimental study of the current limitation process has been conducted using devices of different configurations, which were the combination of high temperature superconducting transformer equipped by the protector against the overloading, that combined the transformer and fault current limiter in the same device [3]. The world-first MgB$_2$-based motor (1.3 kW) from high pressure-high temperature synthesized MgB$_2$ at the ISM, NAS of Ukraine, Kiev, Ukraine, has been developed in cooperation with MAI (Moscow, Russia) and IPHT (Jena, Germany) [5].. The SC rotor of the reluctance electromotor was manufactured from MgB$_2$ synthesized at 2 GPa, 800 $^o$C for 1 h with 10% of Ti addition. The comparative tests of the rotor with MT-YBCO at 20 K have shown that despite the lower SC transition temperature, the efficiency of the electromotor with MgB$_2$ superconductive elements proved to be of the same level as with the MT-YBCO elements at 20 K (Figs. 8b, c). Fig. 8a shows some parts from MgB$_2$-based material produced at ISM, NAS of Ukraine, Kiev, Ukraine, the rotor of the reluctance MgB$_2$-based motor (7). Recently the MgB$_2$ cylinder of outer diameter 21.3 mm, height of 14.1 mm, and wall thickness of 3.5 mm (Fig. 9a) cut from a high-pressure- synthesized sample has been tested in the model of inductive fault current limiter (Physics Department of Ben-Gurion University of the Negev, Israel). The cylinder was placed inside a cooper coil with 1000 turns so that the design represented a two-coil transformer where the cylinder was a single-turn secondary coil. The tested MgB$_2$ cylinder showed high critical current density, the transition from superconductive state to normal and back was very fast (Figs. 9b, c), so it can be used for inductive current fault limiters. The cylinders of larger diameters can provide higher impedance change at the transition from the nominal regime to the limitation regime.

## 6. Conclusions

The developed processes of high pressure-high temperature synthesis, sintering and oxygenation of superconductive ceramic materials allow one to prepare bulk nanostructural materials of Y-Ba-Cu-O and Mg-B-O systems with high SC characteristics, which are very promising for application in cryogenics, especially for cryogenic electrical machines and fault current limiters.

**References.**

**Tables**

Table 1.
Characteristics of MT-YBCO samples after oxygenation in flowing oxygen and at elevated oxygen pressure

| Regime of oxygenation | Oxygen content, x, of $YBa_2Cu_3O_x$ | Vickers micro-hardness, $H_v$, GPa, at a 4.9N-load in ab-plane / ⊥ ab-plane | Fracture toughness, $K_{1c}$, MPa·m$^{0.5}$ | Berkovich nano-hardness, $H_B$, GPa, at a 30 mN-load | Young modulus, E, GPa | Defect density | | |
|---|---|---|---|---|---|---|---|---|
| | | | | | | Macro-cracks, mm$^{-1}$ | Micro-cracks, mm$^{-1}$ | Twins, µm$^{-1}$ |
| Bulk MT-YBCO of Type 1 | | | | | | | | |
| 0.1 MPa, 270 h 440 °C | 6.79 | 4.3±1.1 6.6±0.5 | 0.7±0.2 - | - 9.3±2.0 | - 232±39 | 1.5 | 890 | 0.5-1.7 |
| 16 MPa, 80 h 800 °C | 6.88 | 7.2±0.7 7.5±0.6 | 1.6±0.3 1.95±0.3 | 12.2±0.7 10.0±0.5 | 186±2 217±2 | **0.4** | **200** | **4-11** |
| Thin-walled MT-YBCO of Type 2 | | | | | | | | |
| 0.1 MPa, 270 h 440 °C | 6.79 | 6.8±0.9 7.6±0.1 | 2.45±0.35 1.73±0.13 | 12.6±0.9 10.9±0.5 | 185±9 209±8 | 1.3 | 1050 | 12-16 |
| 16 MPa, 80 h 800 °C | 6.80 | 8.7±0.3 7.6±0.3 | 2.5±0.1 2.8±0.24 | 12.5±0.6 10.3±0.8 | 186±6 218±13 | No cracks | 280 | 20-22 |

**Figure captions**

**Figure 1.** MT-YBCO ceramic blocks of Type 1 (a – as it was grown on seed, b – after grinding of the blocks surface) and of Type 2 (d- as it was grown on seed) and their structure in polarized light before oxygenation (c, e, respectively). The structure of material prepared from Y123 and $Y_2O_3$ ( c ) contains smaller Y211 grains, which are less homogeneously distributed than in the case of the material prepared from Y123 and Y211 (e).

**Figure 2.** The critical current density, $jc$, vs. magnetic field, $\mu oH$, at 77 K for oxygenated MT-YBCO when external field is parallel to the *c*-axis of Y123 (a) and *ab*-plane (b); comparison between the *j*c in both directions of the samples of Types 2 and 1 : curves 1, 2 –thin-walled of type 2 oxygenated under elevated oxygen pressure at 800 °C for 72 h and in flowing oxygen at 440 °C for 270 h, respectively; curves 3, 4, bulk of Type 1 oxygenated under elevated (16 MPa) oxygen pressure and in flowing oxygen, respectively; curves 5 –specially detwined by high pressure (2 GPa) – high temperature treatment Type 1 ceramic with high dislocation density ( about $10^{12}$ $cm^{-2}$)

**Figure 3.** Structure of MT-YBCO after oxygenation (seen in polarized light after etching): under elevated oxygen pressure 800 °C for 80 h of Type 1 (a) and Type 2 (c) and under atmospheric in flowing oxygen at 440 °C for 270 h of Type 1 (b) and Type 2 (d).
TEM images of the MT-YBCO structure after oxygenation under elevated oxygen pressure of two types of ceramics Type 1 (e, f) and Type 2 (g) and bulk (d); structure of MT-YBCO after high pressure-high temperature detwinning (e, h).

**Figure 4.** Comparison of trapped magnetic fields (surface induction) at 77 K after oxygenation under elevated pressure of samples bulk (maximum value 320 mT) (a) and of the thin-wall

geometry (maximum value 664 mT) (b) after field cooling under 2T of MT-YBCO blocks (20 mm in diameter and 20 mm in height) prepared from Y123 and Y211 [26].

**Figure 5**. Structure of $MgB_2$ (obtained by SEM) synthesized at 1050 $^o$C, 2 GPa, 1 h with (a) 10% of SiC (backscattering electron image) and (b) 10% of Ti. – upper and lower right photos are backscattering electron images and lower left photo is secondary electron image.

**Figure 6.** Critical current density, $j_c$, vs. magnetic field for $MgB_2$ (2GPa, 800 and 1050$^o$C, 1h) with 10% of SiC (a) and Ti (b).

**Figure 7.** The structure (obtained by high resolution SEM) of material synthesized from Mg:B = 1:8 (a) and its $j_c$ vs. magnetic field (b), imaginary ($\chi''$) and real ($\chi'$) part of the ac susceptibility (magnetic moment) vs temperature, T(c).

**Figure 8.** (a) Ceramics based on $MgB_2$, synthesized in ISM NAS of Ukraine manufactured under a high pressure of 2 GPa (1-4), by hot pressing (30 MPa), (5, 6), rotor of SC motor with layers manufactured from high-pressure synthesized $MgB_2$ (7); (b, c) characteristics (η - coefficient of efficiency, N - output power and cosφ) at 15-20 K under 210-215 V phase voltage of the SC motor with rotor contained $MgB_2$ layers (b) and MT-YBCO layers (c).

**Figure 9**. (a) ring cut from a high-pressure synthesized sample which has been tested in the model of inductive fault current limiter; (b) typical traces of the voltage across the primary winding (short dash line ) and current (solid line). The arrow shows the moment of the transition. The initial temperature was 4.2 K, the model with the ferromagnetic core; (c) voltage across the primary winding (short dash line) and current (solid line). The transition occurs after several current periods. The initial temperature is 20.3 K, the model without ferromagnetic core.

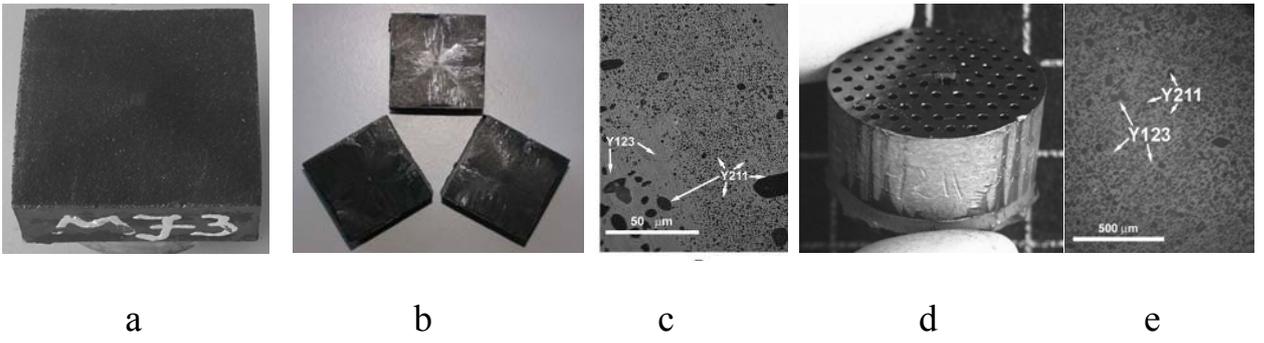

a b c d e

Figure 1.

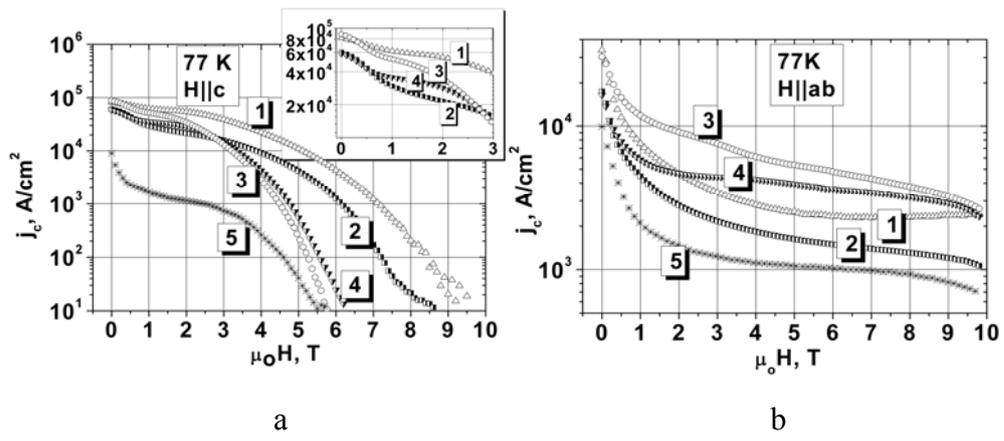

a  b

Figure 2.

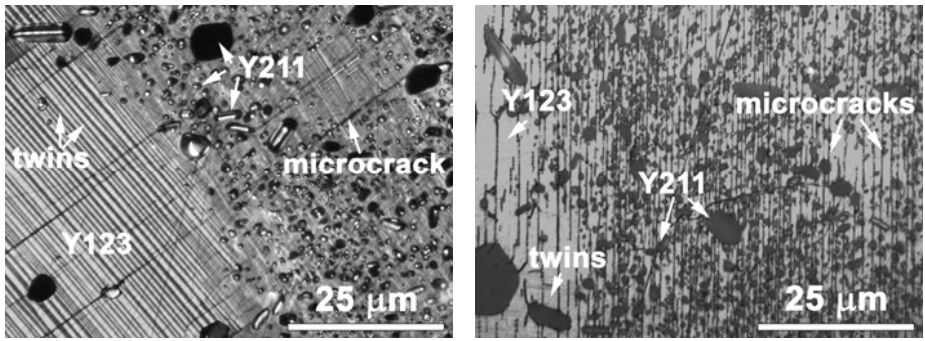
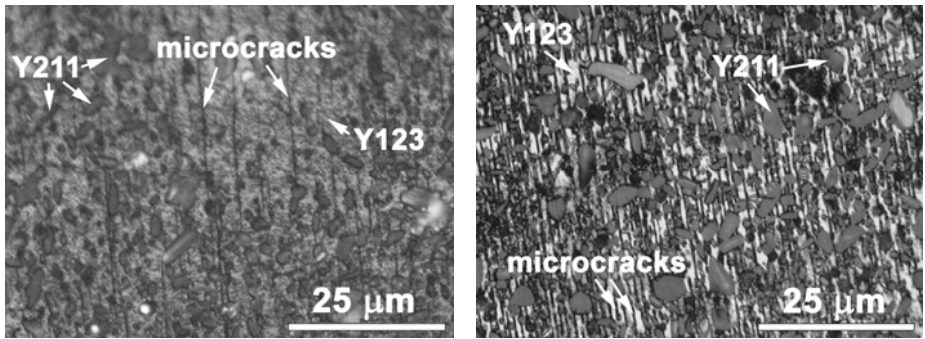
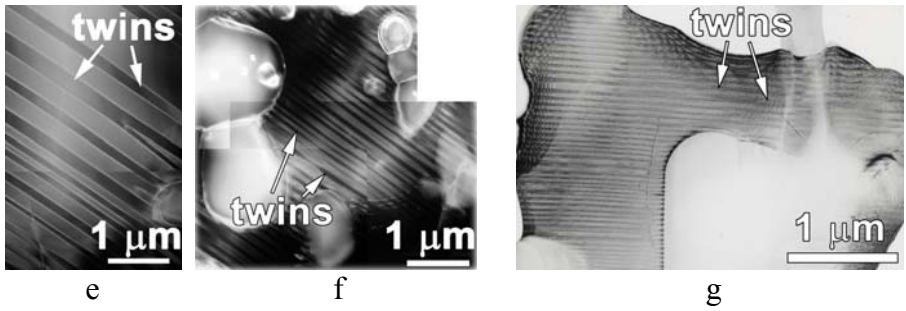
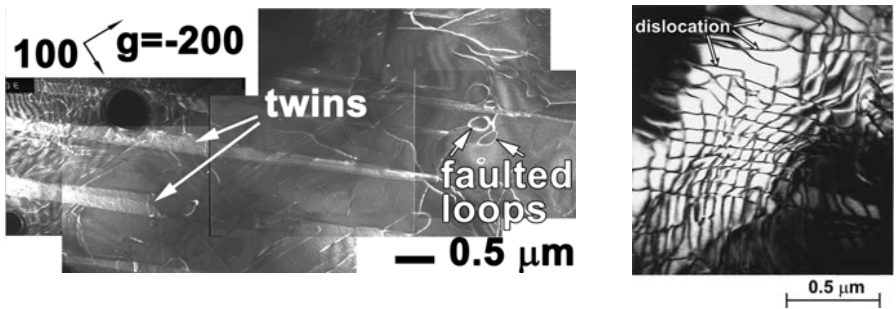

Figure 3.

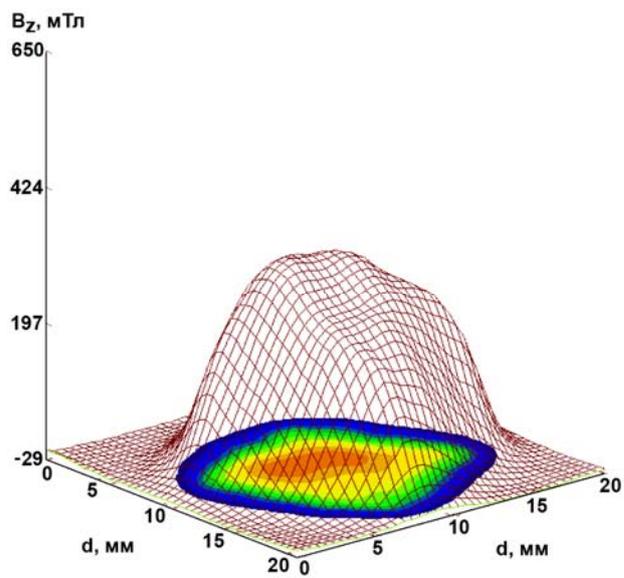 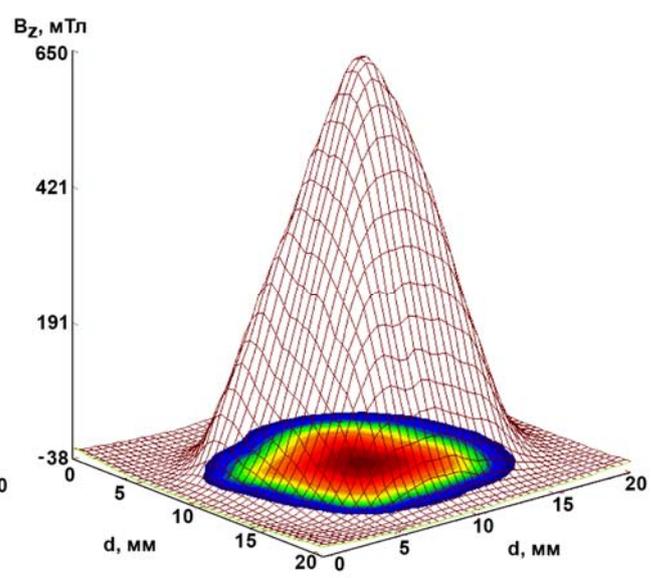

a  b

Figure 4.

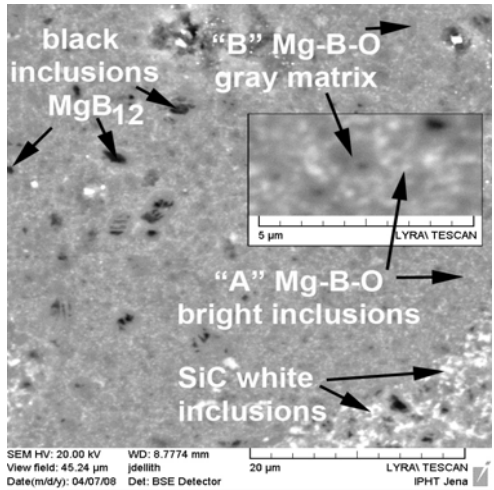 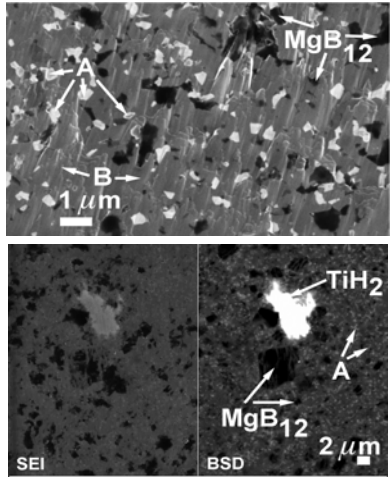

a                                b

Figure 5

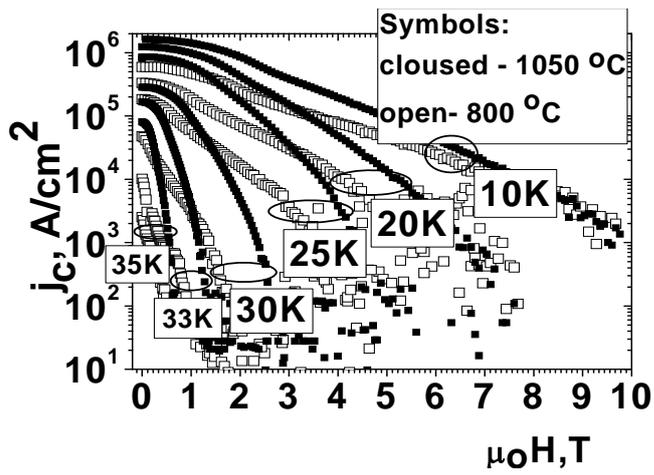

a

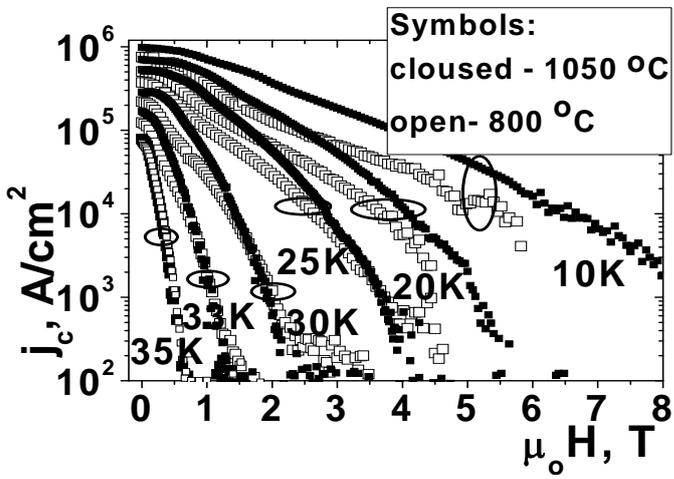

b

Figure 6

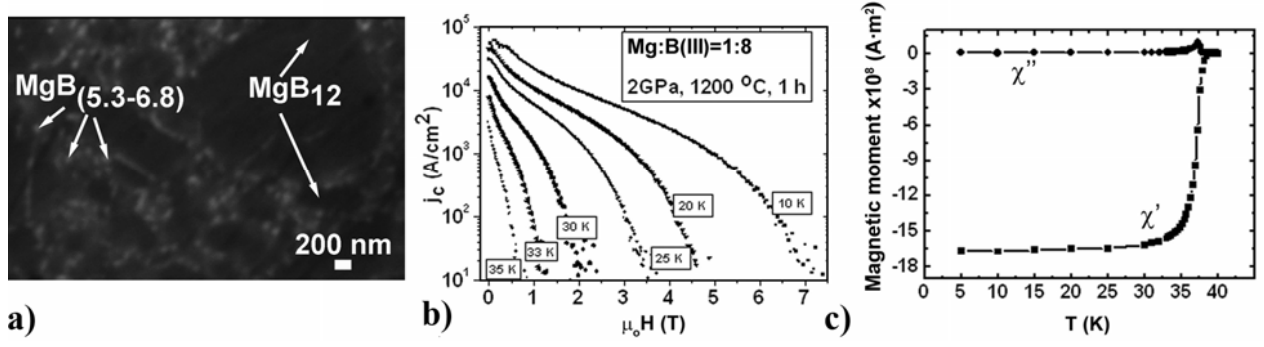

Figure 7.

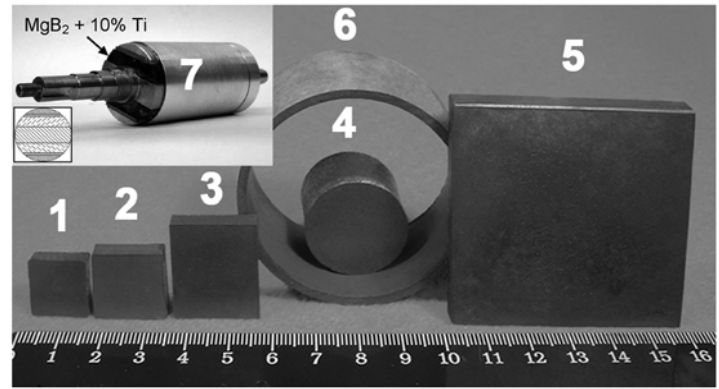
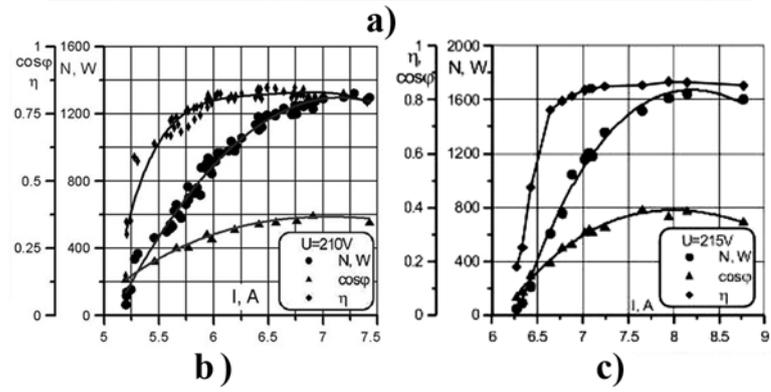

Figure 8

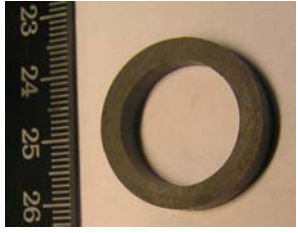 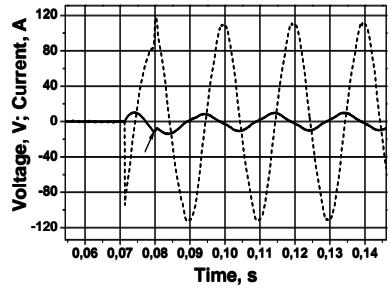 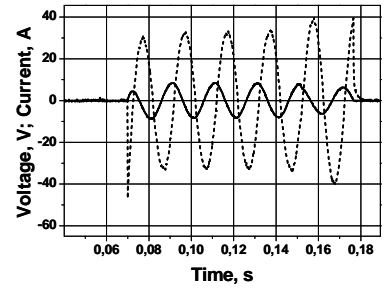

a) b) c)

Figure 9